\documentclass[mathleft
]{an}
\usepackage{graphicx}
\usepackage{times}
\begin{document}

\Pagespan{1}{7}
\Yearpublication{2011}%
\Yearsubmission{2010}%
\Month{XX}%
\Volume{XXX}%
\Issue{XX}%

\title{Magnetic fields of Ap stars as a result of an instability}

\author{R. Arlt\thanks{Corresponding author:
  rarlt@aip.de} \and G. R\"udiger}

\titlerunning{Ap star magnetism as a result of an instability}
\authorrunning{R. Arlt \& G. R\"udiger}
\institute{
Astrophysikalisches Institut Potsdam, An der Sternwarte 16, 
D-14482 Potsdam, Germany
}

\received{30 May 2005}
\accepted{11 Nov 2005}
\publonline{later}


\abstract{%
Ap star magnetism is often attributed to fossil magnetic fields
which have not changed much since the pre-main-sequence epoch
of the stars. Stable magnetic field configurations are known
which could persist probably for the entire main-sequence life
of the star, but they may not show the complexity and diversity
exhibited by the Ap stars observed. We suggest that the Ap 
star magnetism is not a result of stable configurations, 
but is the result of an instability
based on strong toroidal magnetic fields buried in the stars.
The highly nonaxisymmetric remainders of the instability are
reminiscent of the diversity of fields seen on Ap stars. The strengths 
of these remnant magnetic fields is actually between a few per cent
up to considerable fractions of the internal toroidal field; this 
means field strengths of the order of kGauss being compatible with 
what is observed. The magnetic fields emerge at the surface rather
quickly; rough estimates deliver time-scales of the order of a few
years. Since rotation stabilizes the instability, normal A stars
may still host considerable, invisible toroidal magnetic fields.
}
\maketitle

\section{Introduction}
Chemically peculiar stars of types late B, A and early F are 
often accompanied by magnetic fields. We will call them Ap stars
here collectively and will deal with the possible origin of
the magnetic fields observed at the surfaces of these stars.

The observations of magnetic Ap stars show a large diversity
of field strengths, topologies, and rotational periods.
Only a fraction of roughly 10\% of all A-type stars shows the
peculiarities and notable magnetic fields. Ap stars typically
rotate more slowly than normal A~stars. The distributions do
overlap, but the Ap stars form a separate distribution and
are not just a slow tail of the period distribution of normal
A~stars (Abt \& Morrell 1995). 

The magnetic fields have typically strengths of a few kGauss
and are not symmetric with respect to the rotation axis. The
variety of field strengths and geometries is large. Measured
fields are between 0.3~and 30~kG (Donati \& Landstreet 2009).
Especially the very slow rotators among the Ap stars do not 
show any strong magnetic fields of above 7.5~kG (Mathys 2008).
An approximation of the magnetic fields by dipoles leads to an
axis of obliquity against the rotation axis. This was shown
to be large (inclined dipole) for the faster of the Ap stars
and smaller (aligned dipole) for the slower rotators by
Bagnulo et al. (2002). This picture was modified by Mathys (2008)
who found very large obliquity values also for the extremely slowly
rotating Ap stars.

There are also differing results on the evolutionary picture 
of magnetic intermediate-mass stars. Hubrig et al. (2000)
found the Ap star phenomenon to be much less frequent among stars
which have not completed the first 30\% of their main-sequence
life. This was challenged by Landstreet et al. (2007) who
attributed at least part of the effect to the difficulties
in determining the ages of these stars. Young intermediate-mass
stars also show magnetic fields, and the fraction of magnetic
ones of the total number is roughly the same as for Ap stars 
(about 7\%, Wade et al. 2009), indicating that Ap star magnetism
persists from the pre-main-sequence phase into the main sequence. 
In a study relating the magnetic field strength to the age, 
Hubrig et al. (2009) found a decrease of field strength 
with age where the stars were between 0.3 and 14~Myr old, 
concluding that Herbig Ae/Be stars are not the progenitors
of Ap stars.

\section{Ap star magnetism from Tayler instability}
Nearly all magnetic-field configuration are prone to
instability eventually, if the field strength is large 
enough. Magnetic fields $\vec B$ pertaining to electrical currents $\vec j$
will become unstable unless they are force-free, i.e.\ $\vec B || \vec j$
or are balanced by other forces (Duez et al. 2010).
Comprehensive studies of toroidal magnetic fields were
published by Vandakurov (1972) and Tayler (1973). In many 
cases, non-axisymmetric perturbations are the most unstable
ones. The term kink-instability refers to these cases. We
will refer to the whole class of current-driven instabilities
by the term Tayler instability and will not review the extensive 
research that has been done on current-driven instabilities here.

Rotation stabilizes the magnetic fields. A rough estimate
tells that magnetic fields become unstable if $\Omega_{\rm A} \sim
\Omega$, where $\Omega_{\rm A}= B/\sqrt{\mu\rho}\,r \sin\theta$
is the Alfv\'en angular velocity and $\Omega$ is the angular 
velocity of the domain storing the fields (Pitts \& Tayler 1985).
At the expense of longer growth times, smaller fields can 
also become unstable. According to computations by Arlt et al. (2007a)
and Kitchatinov \& R\"udiger (2008), the magnetic fields stored in the
solar tachocline would be limited to a few hundred Gauss. They
can be stabilized further by adding a poloidal magnetic field,
the stability limit then being about 1000~Gauss (Arlt et al. 2007b) 
which still corresponds to fields with $\Omega_{\rm A} \ll \Omega$. 

The stability against non-axisymmetric perturbations is
additionally enhanced if differential rotation is present. 
According to the analysis by R\"udiger \& Kitchatinov (2010),
a weak differential rotation of a few per cent is already
enough to increase the stability limit for the $m=1$ mode
by a factor of three, where $m$ is the azimuthal wave number.
The computations were global in the horizontal direction and
local in the radial direction. We will look at the global 
three-dimensional behaviour in the following study.

Since the magnetic fields of Ap stars are virtually constant
in time, it is interesting to find stable magnetic field
configurations which are not Tayler unstable and could thus
provide the constant fields observed. Braithwaite \& Nordlund
(2006) have computed such equilibria for a non-rotating star.
They are twisted tori in which the poloidal component is the
main contributor to the magnetic energy. More complex structures
of surface magnetic fields were found by Braithwaite (2008).

We are going another way here:
the idea is that the observed surface magnetic fields of
Ap stars are {\em not\/} the manifestation of initially stable magnetic-field
configurations, but that they are the result of unstable
magnetic fields. We explore the possibility that the observed
fields are remnants of the Tayler instability of toroidal 
fields in the stellar interior.

\section{Numerical model}
The simulations employ a spherical shell to mimic the radiative
envelope of an Ap star. The computational domain extends from an
inner radius of $r_{\rm i}=0.5$ to an outer radius of $r_{\rm o}=1$
in normalized units. We need to emphasize though that the simulations
are not meant to cover the very outer zones of the star which are 
characterized by low density and considerably different physics
as compared to the bulk of the purely radiative zone. The system
is simplified to the Boussinesq approximation which ignores
variations of the background density $\rho$ in space and time. It does
allow for small density deviations from the background value thus
permitting also simulations of convection with which we are not
concerned here. The solutions are obtained with the spherical
spectral MHD code by Hollerbach (2000).

The simulations are carried out in non-dimensional units,
where lengths are measured in terms of the stellar radius
$R_\ast$, times in diffusion times, $\tau_{\rm diff}= R_\ast^2 / \eta$
with $\eta$ being the magnetic diffusivity,
and thus velocities and magnetic fields in terms of
$\eta / R_\ast$ and $\sqrt{\mu\rho}\,\eta / R_\ast$, respectively,
where $\mu$ is the magnetic permeability. 
The following non-dimensional equations evolve the
velocity $\vec u$, the magnetic field strength $\vec B$,
and the temperature deviation $\Theta$ from a given
background temperature profile $T_0$:
\begin{eqnarray}
\frac{\partial{\vec u}}{\partial t} &=& 
  -(\vec u\cdot\nabla)\vec u + (\nabla\times \vec B)\times \vec B \nonumber
  +\widetilde{\rm Ra}\,\vec r\,\Theta\\
& &-\nabla p +{\rm Pm}\triangle\vec u,
\label{ns}\\
\frac{\partial{\vec B}}{\partial t} &=&
  \phantom{-}\nabla\times  (\vec u\times \vec B) 
  +\triangle\vec B,
\label{induction}\\
\frac{\partial\Theta}{\partial t} &=&
  -\vec u\cdot\nabla \Theta -\vec u\cdot\nabla T_0 
  + \frac{\rm Pm}{\rm Pr}\triangle \Theta,
\label{temperature}
\end{eqnarray}
where $p$ is the pressure, the Prandtl number Pr is the ratio of the 
viscosity $\nu$ to the thermal diffusivity $\chi$, ${\rm Pr}=\nu/\chi$ 
while the magnetic Prandtl number Pm is the ratio of $\nu$ to the 
magnetic diffusivity $\eta$, ${\rm Pm}=\nu/\eta$. The ratio of 
${\rm Pm}$ to ${\rm Pr}$ is often called the Roberts number $q$. The
density $\rho$ and the permeability $\mu$ are set to unity. The 
background temperature profile follows 
\begin{equation}
  T_0 = \frac{r_{\rm o}r_{\rm i}/r-r_{\rm i}}{r_{\rm o}-r_{\rm i}}
\end{equation} 
accounting for an entirely conductive heat transport with upper and lower
boundary values of 0 and 1, respectively.

The initial velocity field is a differential rotation according to
\begin{equation}
  \Omega(s) = \frac{\rm Rm}{\sqrt{1+s^{\,q}}},
  \label{omega}
\end{equation}
where $s=r\sin\theta$ is the distance from the rotation axis and
Rm is the magnetic Reynolds number defined by
\begin{equation}
  {\rm Rm} = \frac{R_\ast^2 \Omega_\ast}{\eta},
\end{equation}
where $\Omega_\ast$ is the angular velocity of the star. Lines
of constant $\Omega$ are parallel to the rotation axis and
cause the least amount of hydrodynamically induced meridional
flows in which we are not interested here. We assume that the
star has undergone a rotational braking before entering the
main sequence (St\c{e}pie\'n 2000). Since this braking has affected the surface of 
the star, a differential rotation near the ZAMS may be a good model 
for at leatst some of the young intermediate-mass stars. We are
using ${\rm Rm}=20\,000$, $q=4$, and ${\rm Pm}=1$ in all computations.
There is a single simulation employing ${\rm Rm}=40\,000$.

Magnetic fields are measured in terms of Lundquist numbers, which
is the same as the non-dimensional Alfv\'en velocity in our system
of units,
\begin{equation}
  B = \frac{R B_{\rm phys}}{\sqrt{\mu\rho}\, \eta}.
  \label{bconvert}
\end{equation}
The initial magnetic field is a poloidal field of strength $B_0=300$
which is entirely confined in the computational domain. This condition 
is not a requirement for the instability, but it ensures that the 
radial fields finally measured on the surface of the star are
not relics of the initial-field configuration. 

In time-dependent simulations, the magnetic diffusivity $\eta$ is far from 
the stellar one and typically represents a value between the microscopic 
diffusivity of the plasma and the turbulent diffusivity resulting from, e.g.,
averaged convective motions. Of the quantities entering (\ref{bconvert}),
$\eta$ is the one which is known least. It is therefore best to eliminate $\eta$ 
by Rm and thus retrieve the physical magnetic fields by comparing 
its Alfv\'en speed with the rotational velocity,
\begin{equation}
  B_{\rm phys} = \sqrt{\mu\rho}\,\Omega_\ast R_\ast \frac{B}{\rm Rm}.
  \label{bphys}
\end{equation}
The boundary conditions for the flow are stress-free at both
$r_{\rm i}$ and $r_{\rm o}$. There is no imposed velocity, neither
in the bulk of the computational domain nor at the boundaries.
Vacuum conditions are employed at both $r_{\rm i}$ and $r_{\rm o}$
for the magnetic field. Such conditions may look odd at the inner
boundary, but it is a fairly good way of getting the least
amount of artifacts from the inner boundary which, as such, does
not exist in reality. Perfect-conductor conditions are much 
worse, since they cause strong currents near the inner boundary
as soon as the magnetic field tends to fill the whole stellar 
interior from $r=0$ to $r_{\rm o}$ which is prevented by the 
boundary conditions.

Note that the use of spherical harmonics
allows the implementation of exact vacuum conditions which are not
straight-forward in grid codes. The boundary conditions for the temperature
fluctuations are $\Theta(r_{\rm i})=0$ and $\Theta(r_{\rm o})=0$.

\begin{table}
\begin{center}
\caption{The simulations described in this Paper. In the symmetry column,
S means symmetric and A means antisymmetric with respect to the equator.
$B_{\rm top}$ is the dimenionless maximum magnetic field at the surface, 
while $t_{\rm top}-t_{\rm pert}$ is the time when this maximum occurs, measured as a
difference to the instance of perturbation in diffusion times.}
\label{runs}
\begin{tabular}{lccrc}
\hline 
          & {\bf Initial}  &                      &                                   &              \\
{\bf Run} & {\bf symmetry} & $\widetilde{\bf Ra}$ & \multicolumn{1}{c}{$B_{\rm top}$} & $t_{\rm top}-t_{\rm pert}$\\ 
\hline
NLA0      & $\vec u$: S, $\vec B$: A  &    $0$              & 1075  &  $0.00296$ \\
NLS0      & $\vec u$: A, $\vec B$: S  &    $0$              &  167  &  $0.00494$ \\
NLA-1E9   & $\vec u$: S, $\vec B$: A  &    $-1\cdot10^9$    &  199  &  $0.00582$ \\
NLS-1E9   & $\vec u$: A, $\vec B$: S  &    $-1\cdot10^9$    &   64  &  $0.00914$ \\
NLS-5E9   & $\vec u$: A, $\vec B$: S  &    $-5\cdot10^9$    &   60  &  $0.00842$ \\
\hline
\end{tabular} 
\end{center}
\end{table}

The velocity and the magnetic field are expressed by two
scalar potentials each. These and the temperature fluctuations
are decomposed in Chebyshev polynomials in the radial direction
and in spherical harmonics for the horizontal directions. The
potentials are thus functions of the Chebyshev degree $k$, the
Legendre degree $l$, the azimuthal wave number $m$, and the time $t$.
The spectral truncations were at $k_{\rm max}=40$~Chebyshev polynomials
and all spherical harmonics up to $l_{\rm max}=60$ and
$m$ running from $-l$ to $l$.

An implicit scheme integrates the diffusive terms in spectral
decomposition, whereas the nonlinear terms are treated on a
collocation grid in real space. We implemented a variable 
time-step determined by the Courant-Friedrich-Levy (CFL) criterion
from the velocity and the Alfv\'en velocity of the magnetic field. An 
additional safety factor of 0.2 is applied to the maximum possible 
time-step according to the CFL criterion. Since a new time-step 
requires expensive inversions of the time-stepping matrices for 
all five variables, we update the time-step only every 100 integration
steps. This is certainly a compromise but turned out to ensure
-- together with the safety factor of 0.2 -- enough stability 
to run the code through the injection and growth of the 
unstable mode.

\section{Results}
All the simulations are first evolving the axisymmetric initial 
conditions in a three-dimensional domain. The energy in the 
non-axisymmetric modes remains about 30~orders of magnitude
smaller than the energy in the $m=0$ mode. The differential
rotation winds up the initial poloidal magnetic field very
quickly. At the same time, Maxwell stresses grow and diminish
the differential rotation. The whole process reaches a
maximum toroidal field after a time which can be estimated
by $t_{\rm grow}=\sqrt{\mu\rho}R_\ast / B_0$. In our case with 
$B_0 = 300$, this corresponds to a dampig time for the
differential rotation of $t_{\rm damp}=0.0033$ diffusion
times. Note that this time-scale does not depend on the
Reynolds number. In a system of coupled equations of motion
and induction, a stronger differential rotation also means
stronger magnetic fields and thus stronger Maxwell stresses
changing the differential rotation.

\begin{figure}
\begin{center}
 \includegraphics[width=\linewidth]{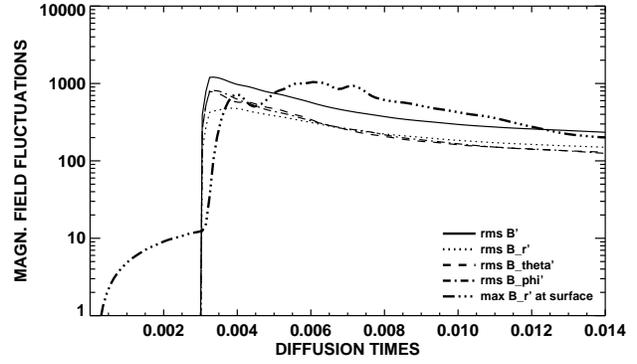} 
 \caption{Magnetic field fluctuations as functions of time
 for the run NLA0 with $\widetilde{\rm Ra}=0$ and a 
 perturbation which is antisymmetric in $\vec B$.}
   \label{spu2b2_nonlin16k}
\end{center}
\end{figure}

\begin{figure}
\begin{center}
 \includegraphics[width=\linewidth]{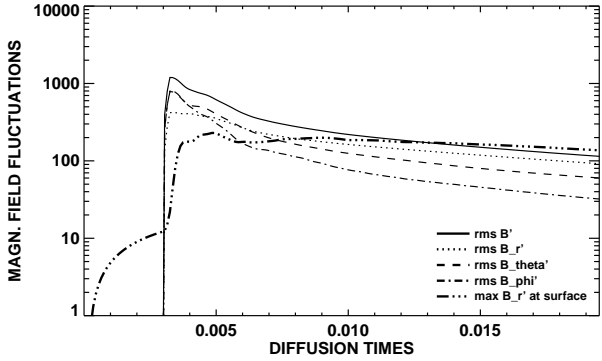} 
 \caption{Magnetic field fluctuations as functions of time
 for the run NLA-1E9 with $\widetilde{\rm Ra}=-10^9$ and a 
 perturbation which is antisymmetric in $\vec B$.}
   \label{spu2b2_nonlin16m}
\end{center}
\end{figure}

An earlier stability analysis delivered the maximum field
strength of the axisymmetric configuration, beyond which
the system becomes unstable under non-axisymmetric perturbations
(Arlt \& R\"udiger 2010). In the present study, we inject a
non-axisymmetric perturbation into the system at $t_{\rm pert}=0.003$ 
diffusion times which is when the axisymmetric configuration is
already supercritical. The perturbation has the topology of a
$P_2^1(\cos\theta)\cos\phi$ spherical harmonic for the antisymmetric
cases (`A' in Table~\ref{runs}) and that of a $P_2^1(\cos\theta)\cos\phi$ 
spherical harmonic for the symmetric cases (`S' in Table~\ref{runs})
in the poloidal potential of the magnetic field. This also corresponds 
to an antisymmetric and a symmetric magnetic field perturbation, respectively. 
The simulations presented here are listed in Table~\ref{runs}.

The nonlinear evolution of the instability of a perturbation
which is antisymmetric in the magnetic field (run NLA0; resulting 
in a symmetric flow perturbation) is shown in 
Fig.~\ref{spu2b2_nonlin16k} for $\widetilde{\rm Ra}=0$ 
in terms of the rms magnetic field components $B_r^{\rm rms}=
\sqrt{\langle B_r'^2\rangle}$ etc. The maximum radial magnetic
field $B_r^{\rm max}$, determined only on the outermost 
surface of the collocation grid, is also shown. The surface
is located at a distance of $\Delta r = 5\cdot10^{-5}$ from the
outer radial boundary. It is interesting to note that the 
maximum surface radial field supercedes the rms values of the internal 
fluctuating magnetic field. The corresponding run NLA-1E9 with 
$\widetilde{\rm Ra}=-10^{9}$ is shown in Fig.~\ref{spu2b2_nonlin16m}.
The emerging fields are lower in general, and the maximum is
reached at a later time. The maximum $B_r$ at the surface 
remains almost constant for the rest of the simulation.

\begin{figure}
\begin{center}
 \includegraphics[width=\linewidth]{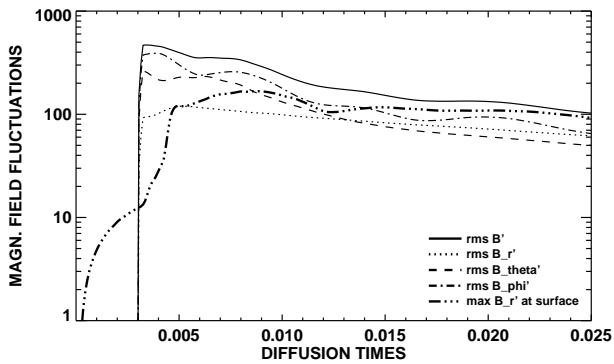} 
 \caption{Magnetic field fluctuations as functions of time
 for the run NLS0 with $\widetilde{\rm Ra}=0$ and a perturbation 
 which is symmetric in $\vec B$.}
   \label{spu2b2_nonlin16a}
\end{center}
\end{figure}

\begin{figure}
\begin{center}
 \includegraphics[width=\linewidth]{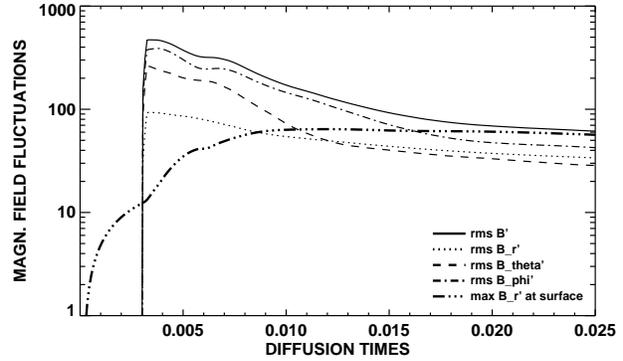} 
 \caption{Magnetic field fluctuations as functions of time
 for the run NLS-1E9 with $\widetilde{\rm Ra}=-10^9$ and a 
 perturbation which is symmetric in $\vec B$.}
   \label{spu2b2_nonlin16f}
\end{center}
\end{figure}

\begin{figure*}
\begin{center}
 \includegraphics[width=15cm]{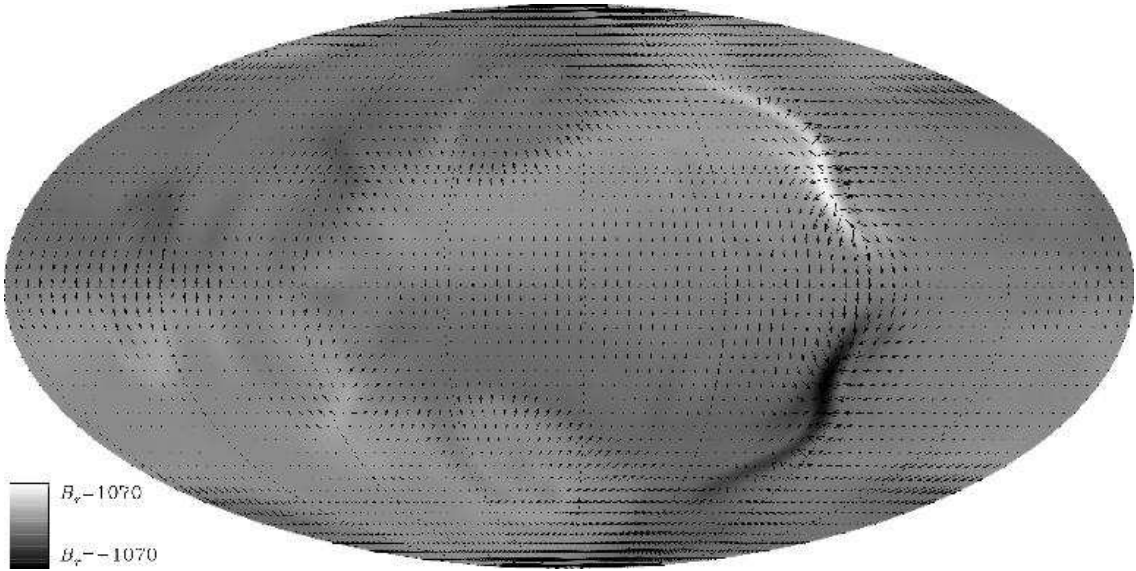} 
 \caption{Surface plot of the magnetic field from the simulation
 NLA0 at $t-t_{\rm pert}=0.00296$. 
 The contours represent $B_r$ while the arrows show $B_\theta$ and 
 $B_\phi$. For the sake of clarity, a smaller number of arrows is plotted
 as compared to the actual number of collocation points in the simulation.}
 \label{nonlin16k}
\end{center}
\end{figure*}

\begin{figure*}
\begin{center}
 \includegraphics[width=15cm]{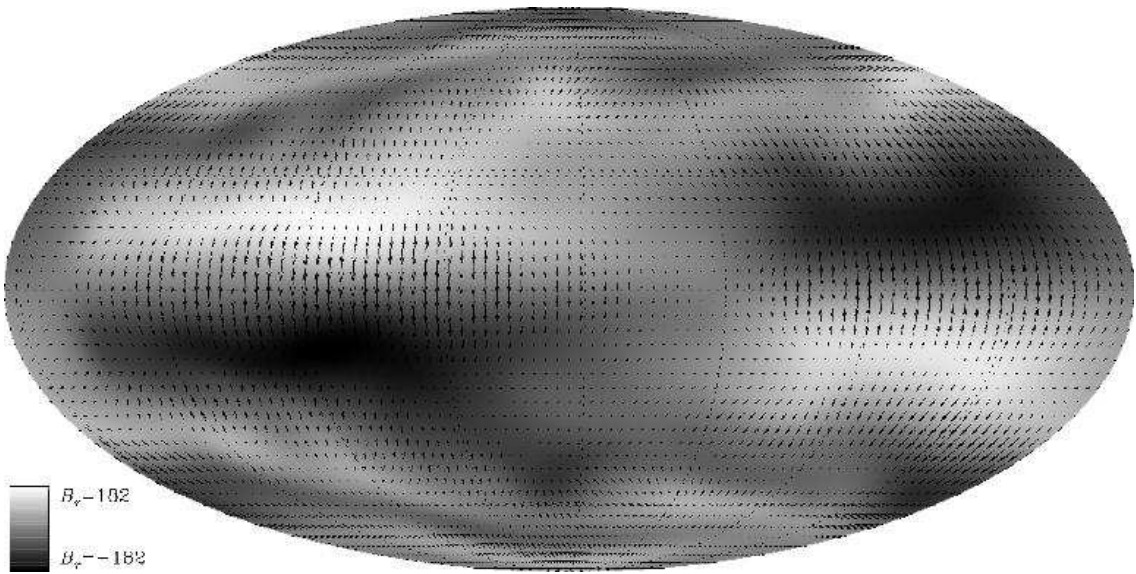} 
 \caption{Surface plot of the magnetic field of the simulation
 NLA-1E9 at $t-t_{\rm pert}=0.00582$. 
 The contours represent $B_r$ while the arrows show $B_\theta$ and 
 $B_\phi$.}
 \label{nonlin16m}
\end{center}
\end{figure*}

Figures~\ref{nonlin16k} and \ref{nonlin16m} show the surface magnetic
fields of the simulations NLA0 and NLA-1E9, respectively. Both are taken
at the moment when the radial surface field reaches its maximum; 
these are at $t-t_{\rm pert}=0.00296$ diffusion times for NLA0 and at 
$t-t_{\rm pert}=0.00582$ for NLA-1E9. The $m=1$ mode is evidently the 
dominating one for NLA-1E9, while the run with $\widetilde{\rm Ra}=0$
shows much smaller azimuthal scales. Stable stratification seems to 
cause smoother surface fields in general, regardless of the symmetry
of the perturbation, and also weaker magnetic fields. A purely 
antisymmetric solution has to show $B_r=0$ in the equatorial plane of 
the rotating sphere, thereby excluding an obliquity of $90^\circ$. It 
is thus interesting to excite a symmetric mode by a symmetric
perturbation, and to test whether maximum obliquity can be achieved.

\begin{figure*}
\begin{center}
 \includegraphics[width=15cm]{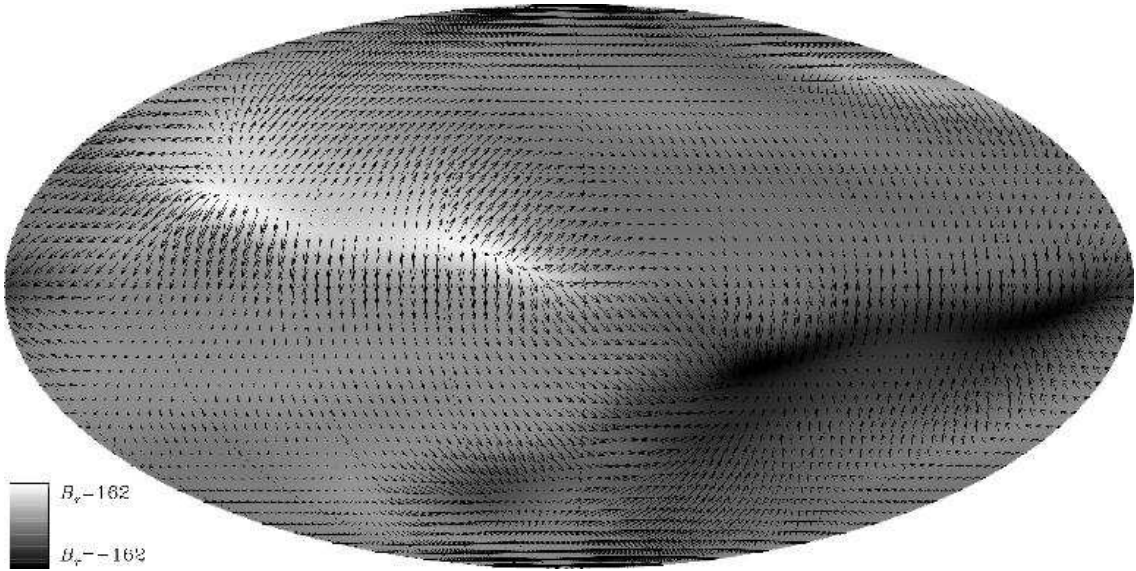} 
 \caption{Surface plot of the magnetic field from the simulation
  NLS0 at $t-t_{\rm pert}=0.00494$. 
  The contours represent $B_r$ while the arrows show $B_\theta$ and 
  $B_\phi$. }
  \label{nonlin16a}
\end{center}
\end{figure*}

\begin{figure*}
\begin{center}
 \includegraphics[width=15cm]{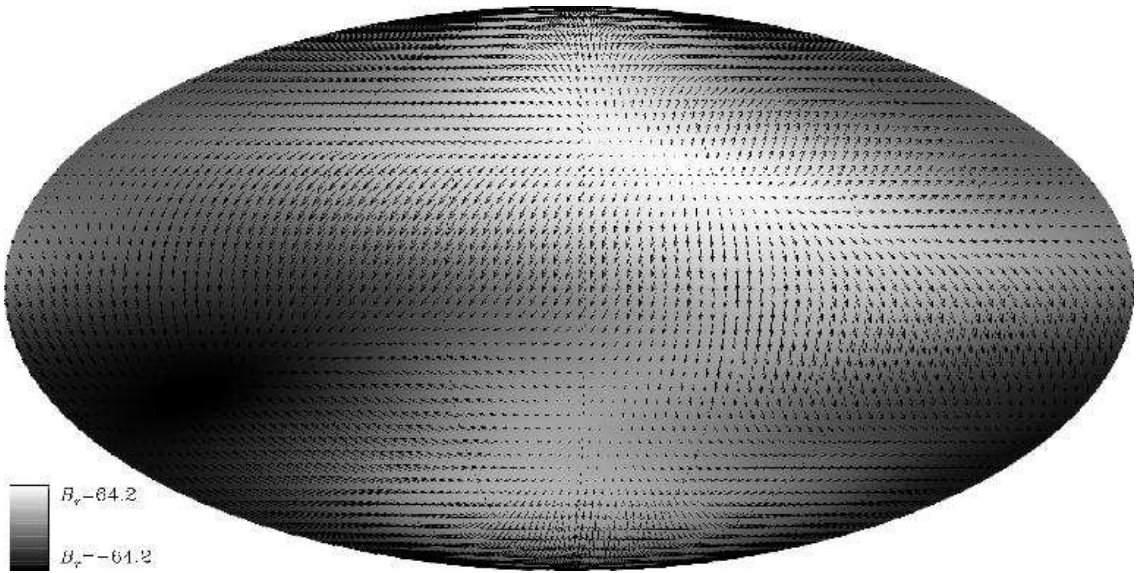} 
 \caption{Surface plot of the magnetic field from the simulation
 NLS-1E9 at $t-t_{\rm pert}=0.00914$. 
 The contours represent $B_r$ while the arrows show $B_\theta$ and 
 $B_\phi$.}
 \label{nonlin16f}
\end{center}
\end{figure*}

Figures~\ref{spu2b2_nonlin16a} and \ref{spu2b2_nonlin16f} show the
corresponding rms magnetic fields for symmetric perturbations. 
The fields emerging are typically weaker and tend to reach their
maximum about 20--30\% later than the fields emerging from an
antisymmetric perturbation. Both symmetries have been shown to
be unstable in the analysis by Arlt \& R\"udiger (2010) near 
$t=0.003$. However, a look at the surface plots in Figs.~\ref{nonlin16a} 
and \ref{nonlin16f} tells us that there is no pure symmetry anymore 
after a certain time; the plots were made at the times of maximum 
surface field, i.e. $t-t_{\rm pert}=0.00494$ diffusion times for NLS0 and 
at $t-t_{\rm pert}=0.00914$ diffusion times for NLS-1E9, respectively.
Numerical noise is apparently growing and delivering a substantial
contribution from the antisymmetric mode. We conclude that
antisymmetric configurations are more likely to become
visible on the stellar surface than symmetric ones. This 
of course excludes an obliquity of precisely $90^\circ$.

The field strengths appearing at the surface are quite substantial.
They range from 4\% to 76\% of the maximum toroidal field strength
inside the computational domain. The corresponding field strengths
are between 1~kG and 29~kG according to (\ref{bphys}) 
for stars with radii between $1.5 R_\odot$ 
and $2.5 R_\odot$ and a rotation period of 10~days. This is a nice 
match with the observed surface magnetic fields. For longer rotation
periods, the dimensionless field strengths correspond to smaller
physical field strengths. 

The next question concerns the time it takes a real star to show
substantial fields on their surfaces, as a result of the Tayler
instability of internal toroidal fields. Questions about
time-scales are always difficult to answer from nonlinear 
simulations since the magnetic Reynolds number will always be
much smaller than the stellar one. This either means that the
angular velocity in the simulation is way too small, or the 
magnetic diffusivity is much larger than the microscopic value
in stars. It is thus necessary to run simulations at various
magnetic Reynolds numbers to see how the results scale with 
Rm. 

While a full exploration of this dependence goes beyond the
scope of this paper, we ran a simulation like NLS0, but with 
${\rm Rm}=40,000$ and obtain a time of maximum surface field of
$t_{\rm top}-t_{\rm pert}=0.00242$. The results indicate 
a $t_{\rm top}-t_{\rm pert} \sim {\rm Rm}^{-1}$
dependence. That has the advantage that the emergence time is simply
a multiple of the angular velocity $\Omega$:
\begin{equation}
  t_{\rm phys} = \frac{C}{{\rm Rm}} \frac{R_\ast^2}{\eta} 
  = \frac{C}{\Omega},
\end{equation}
where $C=98$ is the result of fitting the two points with a 
power law. A 10-day rotation period of the star results in an
emergence after $0.43$~yr for NLS0, while it is ten times longer for a
100-day rotation period. The longest emergence delays seen in
the simulations are about 8~yr with a strongly stable
stratification.

\section{Discussion}
We simulated the nonlinear, three-dimensional evolution of the 
Tayler instability in a spherical shell. The instability 
provides magnetic field configurations of mostly large scales 
with a preference to modes which are nonaxisymmetric and antisymmetric
with respect to the equator. The field strength of the maximum
radial magnetic field at the stellar surface was found to be between
1~and 29~kG which is what is observed on most Ap stars. The emergence 
of the remnant fields from the instability at the surface of the star 
is delayed by several tens of stellar rotations. This estimate 
holds only if  $t_{\rm top}-t_{\rm pert} \sim \Omega^{-1}$, however. 
In reality, it is a lower limit for the emergence delays.

This is certainly a too simple way of getting the time delays
between the instability and flux emergence at the surface, but
we can conclude that the time necessary to observe the remnant 
fields are shorter than evolutionary time-scales. By contrast,
Arlt \& R\"udiger (2010) argued that the flux rise is of diffusive
nature and obtained delays of several Myr. The present study indicates 
that the time-scale will be shorter. As a consequence, if magnetic 
fields would appear on Ap stars at later times during their evolution,
it must be by other factors which influence the stability limits
of the potentially stored toroidal fields. Since the instability
as well as the rise of fields to the surface are very fast, it
is highly unlikely that the phenomenon itself is seen in progress. 

As the instability drains energy from the magnetic field,
the growth quickly halts and the remaining magnetic field suffers
only from the flows excited by the instability and from Ohmic 
diffusion which is very slow. The rms velocities are about 
three times weaker than the rms magnetic fields shown in 
Figs.~\ref{spu2b2_nonlin16k}--\ref{spu2b2_nonlin16f} and do not 
cause quick changes of the surface structure. This gives the 
impression of stationary magnetic fields, while the period
of instability is very short and most likely missed by observations.

The finding by Mathys (2008) that the very slow rotators are not
hosting any fields above 7.5~kG would be compatible with the fact
that an originally slower rotation of the pre-main-sequence star
has not allowed very strong toroidal fields to build up, since the
instability limit is lower for slow rotation, whence the smaller 
remnant fields from the Tayler instability.

The drawback of the present approach is the Boussinesq 
approximation which is actually valid for small deviations
from the adiabatic temperature gradient. The imitation of
a very stable stratification by a highly negative $\widetilde{\rm Ra}$ 
is still telling us qualitative features of the emergence of surface 
fields, but quantitatively, we need to be careful with emergence 
times and flux. The same also holds for computations in the anelastic 
approximation. First computations of the scenario in a fully compressible 
spherical shell are on the way.

Normal A stars would thus still be hosting considerable toroidal
fields. The emergence times of stable magnetic-field configurations
are very long and probably beyond 100~Myr (Mestel \& Moss 2010).
Since they rotate typically faster than Ap stars, normal A stars
may have a higher threshold for the Tayler instability and could
thus be able to keep strong toroidal magnetic fields in their interiors
without showing substantial fields on the surface. The implication
would be, however, that as soon as the stars start to become giants,
their radii grow, and the rotation periods increase substantially. 
The fields will no longer be stabilized and must become unstable.
This would imply that nearly all stars on the giant branch having
intermediate-mass stars as progenitors should show magnetic fields.
These are of course much weaker because of the larger radius
and the steep decrease of field strength with radius, especially
for higher modes than dipoles. The giant EK~Eri has been considered
a descendant of an Ap star (St\c{e}pie\'n 1993; Dall et al. 2010),
i.e. the star would have shown surface magnetic fields through nearly
all its life. However, the magnetic fields may have been emerging only
when the star evolved away from the main sequence and became a slow
rotator, and the progenitor would actually be a normal A~star. Since the
star has a relatively low mass among the ``A-star descendants'' its
age of roughly 1~Gyr may be even compatible with the diffusive emergence 
of fields discussed by Mestel \& Moss (2010) though.

The critical question now is how Ap stars are discriminated from
normal A stars in an early stage of stellar evolution. This problem
cannot be solved in the context of this Paper, but it is suggested 
that it is differences in the rotational evolution during the
pre-main-sequence phase that decides whether stars evolve into
normal A stars or Ap stars. It would not be necessary to think
of a presence or absence of magnetic fields during star formation.
These ``primordial'' fields are most likely processed during the 
Hayashi phase and will be highly modified.

\end{document}